# Higher-order anisotropic flow correlations in Xe–Xe collisions at $\sqrt{s_{NN}} = 5.44$ TeV

Saraswati Pandey[1,a] , B. K. Singh[1,2,b]

[1] Department of Physics, Institute of Science, Banaras Hindu University (BHU), Varanasi 221005, India
[2] Discipline of Natural Sciences, PDPM Indian Institute of Information Technology Design and Manufacturing, Jabalpur 482005, India

**Abstract** By employing Monte Carlo HYDJET++ model (HYDrodynamics plus JETs), we produce anisotropic harmonic flow coefficients $v_n$ ($n = 4$–$7$) in deformed Xe–Xe collisions at $\sqrt{s_{NN}} = 5.44$ TeV. We measure these harmonics with respect to a plane constructed using lower-order Fourier harmonics $v_2$ and $v_3$ (produced using reaction plane method). The cross-talk of elliptic and triangular flows in the model generates both even and odd harmonics of higher order. By combining analyses of higher harmonics with analyses of $v_2$ and $v_3$, one can eliminate the uncertainty in modeling anisotropic flow from initial conditions and define quantities that only involve nonlinear hydrodynamic response coefficients. In this process, we study the individual response of higher-order flow coefficients to the lower-order flow coefficients through a power-law (relation $v_n/v_m^{n/m}$) scaling technique as a function of collision centrality. We report that these higher-order flow coefficients $v_n$ ($n = 4$–$7$) are centrality dependent and strongly correlated with elliptic and triangular flow. The results are compared with data from recent ALICE, ATLAS and CMS experiments at LHC.

## 1 Introduction

In the past few years or so, RHIC (Relativistic Heavy-Ion Collider) and LHC (Large Hadron Collider) experimental programs have explored anisotropic flow [1] and its associated fluctuations [2–4] miraculously to a remarkable degree of precision [5–8]. These studies essentially included detailed analyses of the higher Fourier harmonics $v_4$, $v_5$, $v_6$ and their correlations with the lower Fourier coefficients $v_2$, $v_3$. The prime goal of the presented study is to show that higher-order Fourier harmonics can be expressed as the combination of lower Fourier harmonics under Monte Carlo HYDJET++ framework and examine the outcomes by comparing the results from LHC experimental data (ALICE, ATLAS, etc.). The comparison of these higher-order Fourier harmonics from either hydrodynamical calculations [9–11] or various transport models [12] offers a little insight into the underlying physics involved in them.

Theoretical speculations strongly depend on the initial density profile of the model. This is the main source of ambiguity in modeling anisotropic flow [13]. The physics of higher-order Fourier harmonics should not be so complicated. For example, in an ideal hydrodynamics, at high transverse momentum, the ratio of $v_4/(v_2)^2$ is $1/2$ [14]. This means that the higher-order azimuthal Fourier harmonics (n>3) emanate not only from their respective linear eccentricities $\epsilon_n$ but are also contributed by the nonlinear response terms (nonlinear mixing of lower-order harmonics $v_2$ and $v_3$) [15–18] from $v_2$ and $v_3$ that become visible at high $p_T$. However, from article [19], it was argued that the nonlinear response coefficients are independent of the initial density profile in a given centrality class of Pb–Pb collisions at 2.76 TeV, where Pb nucleus is considered to be spherically symmetric. So, working in deformed collision systems would offer a deeper insight to this speculation where the initial density profile is altered and thereby may affect anisotropic flow. In another work on similar collision system [20], it was verified that the correlations between the leading linear terms $v_{nL}$ along with all mode coupling terms $\langle v_{nL}\rangle$ could be ignored. Hence, one may approach in the way to study the part of flow harmonics arising solely from significant nonlinear part of lower-order flow coefficients.

Centrality dependence of $v_n - v_m$ correlations is made up of linear and nonlinear parts which showed a suitable match with the experiment qualitatively [21]. It is claimed that the nonlinear portion of the correlation elicited by different methods encompasses nonlinearities not only from initial state (via nonlinear eccentricity correlations) but also from hydrodynamic evolution (via mode-coupling effects) which cannot be separated model-independently [20, 21]. These two kinds of nonlinearities are primarily induced by the elliptic geometric deformation of the nuclear overlap region in non-central A+A collisions. This can be attested both theoretically and experimentally in ultra-central U+U (uranium isotope is a deformed nucleus) collisions [21], as in such collision systems, the nuclear overlap region is elliptically deformed even at zero impact parameter. Here, it is preferred to perform such study on much smaller, Xe+Xe collision system but at much higher center-of-mass energies (LHC energies).

[a] e-mail: saraswati.pandey13@bhu.ac.in
[b] e-mails: bksingh@bhu.ac.in; director@iiitdmj.ac.in (corresponding author)

Springer

Recent work on Xe–Xe collisions under Monte Carlo HYDJET++ model [22, 23] presented a rigorous study of flow harmonics as a function of transverse momentum and collision centrality. Correlation between the flow coefficients was found to be dependent on geometry and on the system-size of collision. Hexagonal flow $v_6$ in Pb–Pb collisions at 2.76 TeV using HYDJET++ model was studied recently [24] where, nonlinear contributions from $v_2$ and $v_3$ to $v_6$ were studied as a function of centrality and transverse momentum. In this article, the analysis of anisotropic flow harmonics is extended to much higher harmonic (up to seventh) order. We will investigate the response of these higher-order Fourier coefficients to the lower-order flow harmonics with respect to collision centrality through power-law scaling techniques. In Sec. 2, we will briefly discuss the formulation of HYDJET++ model. The results and discussions of the measurements will be debated along with the comparison with available experimental data at LHC energies in Sec. 3. Lastly, we will present the summary of our study in Sec. 4.

## 2 Model formalism

HYDJET++ (HYDrodynamics plus JETs) is a Monte Carlo event generator designed to simulate relativistic heavy-ion collisions. The model works by superimposing the soft hydro-type state and the hard state resulting from multiparton fragmentation, simultaneously treating both the states independently. It gives an exhaustive approach to the soft hadroproduction (collective flow effects and resonance decays) and also to the hard parton production along with the known medium effects (jet quenching, nuclear shadowing, etc.). The in-depth details of the model and the procedure of simulation can be found in the corresponding articles [25, 26] and the references there within. A crisp and succinct glimpse of the model is as follows:

The soft part of a HYDJET++ event is a thermal hadronic state produced on chemical and thermal freeze-out hypersurfaces derived from a parameterization of relativistic hydrodynamics with preset freeze-out conditions [27, 28]. The hadronic matter created in a nuclear collision reaches a local equilibrium after a short period of time (< 1 fm/c) and then expands hydrodynamically. In HYDJET++, a scenario of different chemical and thermal freeze-outs ($T_{ch} \geq T_{th}$) is used. The system expands hydrodynamically with frozen chemical composition in between these two freeze-outs, then cools down and the hadrons stream freely as soon as $T_{th}$ is reached. The concept of single freeze-out is eliminated as the particle densities at the chemical freeze-out stage are too high to consider the particles as free streaming [29].

The model for hard part of a HYDJET++ event is similar to that of the HYDJET event generator [30–32]. The hard state is treated using Pythia Quenched (PYQUEN) model [30]. PYQUEN model modifies a jet event generated by PYTHIA by producing nucleonic collision vertices according to Glauber model at a certain impact parameter. PYTHIA is an event generator that simulates hard nucleon–nucleon (NN) collision. It considers only those events whose generated total transverse momentum is higher than $p_T^{min}$. Here $p_T^{min}$ is an important free parameter in HYDJET++ which separates soft part of the event from the hard part. It is an input parameter of the model and not the final state quantity output by the model. $p_T^{min}$ (= 10.5 GeV/c) is the minimum transverse momentum transfer of hard parton–parton scatterings in GeV/c [25]. In the HYDJET++ framework, partons produced in (semi-)hard processes with the momentum transfer lower than $p_T^{min}$ are thermalized. So their hadronization products are included in the soft part of the event automatically. In the model framework, this $p_T^{min}$ parameter is used to calculate the mean number of jets produced in AA events at a given impact parameter b. Further, the process takes place by rescattering-by-rescattering simulation of the parton path in the dense zone and the radiative and collision energy losses [33–37] associated in there. Then, final hadronization is carried out using Lund String Model [38] for hard partons and in medium emitted gluons. An impact parameter-dependent parameterization under the framework of Glauber–Gribov theory is used to incorporate the known medium effect such as nuclear shadowing [39, 40].

### 2.1 Anisotropic flow $v_n$ in HYDJET++

Anisotropic flow describes a collectivity among particles produced in heavy-ion collision. It is perceived as an important observable which furnishes key information on the early time evolution of the nuclei interaction. It is usually measured by the Fourier harmonics with respect to the reaction plane [41, 42] as:

$$v_n \equiv \langle \cos[n(\psi - \psi_R)] \rangle \tag{1}$$

where $\psi$ = azimuthal angle of the produced particle,

$n$ = harmonic value, and $\psi_R$ = orientation of the reaction plane, and angular brackets denote an average over all particles belonging to some phase-space region and over many events.

For $n = 1$, it is called Directed flow. HYDJET++ does not consider directed flow $v_1$ of particles, which is essentially zero and much weaker than the $v_2$ and $v_3$ [43].

For $n = 2$, it is called elliptic flow. In HYDJET++ framework, the reaction plane of order two is zero for all the events. $v_2$ in terms of particle momenta is characterized by-

$$v_2 = \left\langle \frac{p_x^2 - p_y^2}{p_x^2 + p_y^2} \right\rangle = \left\langle \frac{p_x^2 - p_y^2}{p_T^2} \right\rangle \tag{2}$$

However, in HYDJET++ framework, this elliptic flow coefficient is calculated using the parameters $\epsilon_2(b)$ and $\delta_2(b)$ known as the spatial anisotropy and momentum anisotropy, respectively, through Eq. (1). $\epsilon_2(b)$ exemplifies the elliptic modulation of the final freeze-out hypersurface at a given impact parameter b, whereas $\delta_2(b)$ deals with the alteration of flow velocity profile. These coefficients are impact parameter dependent. HYDJET++ model reports the mid-rapidity area of heavy-ion collisions rather than the fragmentation regions (high-rapidity regions). HYDJET++ has no evolution stage; as a result, it cannot trace, for instance, the propagation of energy and density fluctuations of the initial state. Therefore, it only handles the final components of the anisotropic flow.

Elliptic flow is produced by the contribution of soft hadrons with low value of transverse momentum while hadrons having high value of transverse momentum are restrained. Soft particle emission in HYDJET++ is produced from a freeze-out hypersurface at the time of freeze-out. As a result, $v_2$ is not directly related to the initial spatial anisotropy ($\epsilon_0$) of the participating nucleons like it happens in AMPT model. In HYDJET++ model, the fireball produced has geometrical irregularities in different directions of phase space at the time of freeze-out. These irregularities are indirectly related to the initial spatial distribution of the participating nucleons in the collision region. The transverse radius $R_{ell}(b, \phi)$ of the fireball formed upon collision in the given azimuthal direction $\phi$ is related to spatial anisotropy at the time of freeze-out by:

$$R_{ell}(b,\phi) = R_f(b)\frac{\sqrt{1-\epsilon_2^2(b)}}{1+\epsilon_2(b)\cos 2\phi}, \quad (3)$$

where

$$R_f(b) = R_0\sqrt{1-\epsilon_2(b)}. \quad (4)$$

Here $R_0$ is the freeze-out transverse radius in absolute central collision with b = 0. For triangular flow $v_3$ in HYDJET++, the model has parameter $\epsilon_3(b)$ for spatial triangularity of the fireball. The modified radius of the freeze-out hypersurface in azimuthal plane is expressed as:

$$R(b,\phi) = R_{ell}(b)\{1 + \epsilon_3(b)\cos[3(\phi - \psi_3^{RP})] + \cdots\}. \quad (5)$$

where $\phi$ = spatial azimuthal angle of the fluid element relative to the direction of the impact parameter.

The phase $\psi_3^{RP}$ introduces the third harmonic having its own reaction plane, distributed randomly with respect to the direction of the impact parameter ($\psi_2^{RP} = 0$). This new anisotropy parameter $\epsilon_3(b)$ is handled independently for each centrality or dependent using $\epsilon_0(b) = b/2R_A$ where $R_A$ is the nucleus radius. Such modifications do not affect the elliptic flow (controlled by $\epsilon_2(b)$ and $\delta_2(b)$). Hence, the triangular dynamical anisotropy can be incorporated by the parameterization of the maximal transverse flow rapidity [23],

$$\rho_u^{max}(b) = \rho_u^{max}(0)\{1 + \rho_{3u}(b)\cos[3(\phi - \psi_3^{RP})] + \cdots\}. \quad (6)$$

Similarly, the maximal transverse flow rapidity [25] after the parameterization of the four-velocity $u$ up to the seventh-order harmonics is given as,

$$\rho_u^{max}(b) = \rho_u^{max}(0)\{1 + \rho_{3u}(b)\cos 3\phi + \rho_{4u}(b)\cos 4\phi \\ + \rho_{5u}(b)\cos 5\phi + \rho_{6u}(b)\cos 6\phi + \rho_{7u}(b)\cos 7\phi\}. \quad (7)$$

Hence, we can calculate higher harmonics with respect to the direction of the impact parameter b. Again, these new anisotropy determiners $\rho_{3u}(b)$, $\rho_{4u}(b)$, ... can be treated both independently as well as being dependent via initial ellipticity $\epsilon_0(b) = b/2R_A$. Here, we have treated the parameters $\rho_{3u}(b)$ and $\rho_{4u}(b)$ independently and varied them with centrality while the higher-order anisotropy determiners are incorporated through the nonlinear mode-mixing. The whole description of the flow calculation within HYDJET++ framework can be found in detail within the references [23, 25]. Actually, anisotropic flow solely results from the $\phi$ of the fluid velocity. This is performed by independently treating the anisotropy determiners which is the transverse flow rapidity parameter corresponding to each harmonic. In this way, we mimic the eccentricity scaling behavior for each flow harmonic independently. However, this is only performed for $v_3$ and $v_4$. For higher harmonics, the idea of nonlinear mode-mixing is incorporated similar to the one performed at LHC energies for spherical collision systems [24]. While the simulation happens, plain or normal averaging is performed over all particles in each event and over all events.

Elliptic flow $v_2$ of particles contributes to all even harmonics, i.e., $v_4$, $v_6$, etc. For example, quadrangular flow $v_4$ in HYDJET++ is determined by the elliptic flow of particles, governed by hydrodynamics, and the particles coming from jets [44]. The coupling between the elliptic and triangular flows results in the appearance of higher odd harmonics $v_5$, $v_7$, and so on in the model. Similar to $v_2$, triangular flow $v_3$ should also individually contribute to $v_6$, $v_9$, etc. The reaction plane of second order $\psi_2$ being zero for each event has the new phase $\psi_3$ randomly distributed with respect to its position of $\psi_2$. This implies that, integrated $v_3$ measured in the $\psi_2$ plane is zero. Therefore, the produced $v_3$ arises from its own independent reaction plane. However, in case of higher-order flow harmonics $v_n(n \geq 4)$ the situation is different. $v_2$ and $v_3$ meddle with each other generating both even and odd higher azimuthal harmonics, $v_4$, $v_5$, $v_6$, etc., called as overtones. This is because in HYDJET++ these higher Fourier harmonics do not generate their own intrinsic event planes $\psi_n^{RP}$. For instance, $v_5$ is generated by the interference between elliptic and triangular flows ($v_5 \propto v_2 \times v_3$).

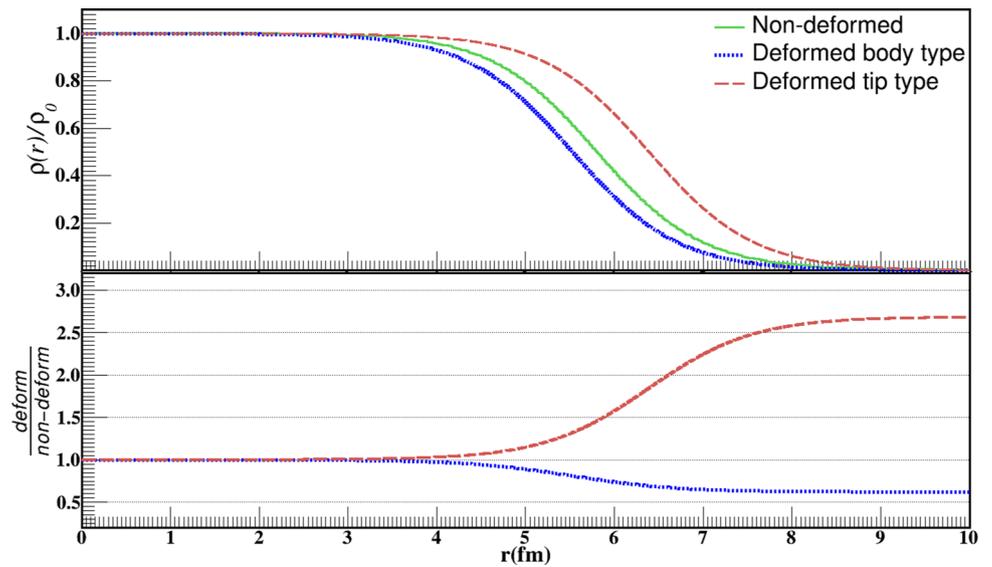

**Fig. 1** The nuclear density profile for Xenon nucleus. Shown are the non-deformed and deformed Woods-Saxon nuclear density profiles. Two types of geometries are shown: body-type and tip-type. The bottom panel shows the ratio of the density profile with nuclear deformation in these two geometries with respect to zero deformation

Hexagonal flow $v_6$ emerges by the nonlinear interplay between elliptic and triangular flow $v_6 \propto (v_2^3 + v_3^2)$. Thus, higher azimuthal flow harmonics emanate in HYDJET++ model solely due to the existence of $v_2$ and $v_3$ through the interference between them. The above description of $v_n (n \geq 4)$ is the focussed part of our work, wherein we will examine the response of the flow harmonics as a function of collision centrality along with the examination of the contributions of the lower-order flow coefficients $v_2$ and $v_3$ to them through $v_n/v_m^{n/m}$ scaling and correlation techniques.

## 2.2 Intrinsic deformation in Xenon nucleus

HYDJET++ model does not include nuclear density profiles for nuclei with some intrinsic deformation. Therefore, the nuclear density function needs to be modified for deformed nucleus like U, Xe, etc. In our present work, we have used Modified Woods-Saxon (MWS) to calculate initial distributions of partons, etc., for body, tip or other random configuration collisions of Xenon nuclei. The MWS nuclear density profile function for deformed nuclei is expressed as:

$$\rho(r, z, \theta) = \frac{\rho_0}{1 + \exp\frac{(r - R(1 + \beta_2 Y_{20} + \beta_4 Y_{40}))}{a}} \quad (8)$$

where $\rho_0 = \rho_0^{const} + \text{correction}$, $\rho_0^{const} = \frac{M}{V} = \frac{3A}{4\pi R_A^3}$, $R_A = R(1 + \beta_2 Y_{20} + \beta_4 Y_{40})$, $R = R_0 A^{1/3}$, where $R_0 = 1.15$fm. The correction term is calculated as $= \rho_0^{const} \times (\pi f/R_A)^2$, where f = 0.54 fm, $\beta_2 = 0.162$ and $\beta_4 = -0.003$ are the deformation parameters, a = 0.59 fm is the diffuseness parameter, $Y_{20} = \sqrt{\frac{5}{16\pi}}(3\cos^2\theta - 1)$, and $Y_{40} = \frac{3}{16\sqrt{\pi}}(35\cos^4\theta - 30\cos^2\theta + 3)$ are the spherical harmonics.

The values of the deformation parameters have been taken from the article [48]. However, HYDJET++ model works in cylindrical coordinate system; therefore, the modified nuclear density profile function is converted from spherical polar coordinates $(r, \theta, \phi)$ to cylindrical polar coordinates $(\rho, z, \psi)$. The effect of deformation in the nuclear density profile for Xenon nucleus can be visualized in Fig. 1 where the upper panel shows the normalized nuclear density profile of Xenon with and without deformations. Here, two types of geometries are shown: body-type and tip-type. The geometrical configuration is mainly controlled by $\theta$ and all other coordinates integrated over the same range. The lower panel shows the ratio of the density profile with nuclear deformation in these two geometries, with respect to zero deformation. The collision systems formed from deformed nuclei produce a medium having anisotropic shape in the transverse plane with intrinsic elliptic deformation. Introducing the deformation parameter $\beta$ with some magnitude in the nuclear density profile makes the medium with some initial eccentricity $\epsilon$ contributed by the quadrupole deformation of colliding nuclei. In this way, the deformation affects the flow harmonics.

## 3 Results and discussion

We have generated $10^6$ events using the modified HYDJET++ model in each of seven centrality classes for minimum bias Xe–Xe collisions at 5.44 TeV center-of-mass energy. Primary charged particles with $|\eta| < 0.8$ with $p_T > 0.2$ GeV/c, and $|\eta| < 2.5$ with $0.5 < p_T < 60$ GeV/c have been considered separately. It was demonstrated that tuned HYDJET++ model can reproduce LHC data on centrality and transverse momentum dependence of charged particle multiplicity density, transverse momentum $p_T$ and anisotropic flow $v_n$(n $\leq$ 4) spectra up to $p_T \sim 5.0$ GeV/c and (0–60)% centrality range [22, 23]. In the present study, we have compared HYDJET++ simulation results with the LHC experimental data from all the three detectors (ALICE, CMS, and ATLAS) [45–47] for our analysis. The data measured from ALICE and CMS experiments have been calculated using 2-particle cumulant technique, whereas the ATLAS experimental data have been measured using Scalar Product Method, the average value of each flow coefficient being root mean square value. HYDJET++ model results have been presented from the basic methodology of the model in which we calculate the lower-order flow harmonics $v_2$ and $v_3$ using the reaction plane method while the higher orders via nonlinear mode-mixing. We study $v_n$-spectra of primary charged hadrons as a function of collision centrality in the proposed kinematic ranges. We will also investigate the response of higher-order flow coefficients to the lower-order flow coefficients through power-law (relation $v_n/v_m^{n/m}$) scaling technique as a function of collision centrality.

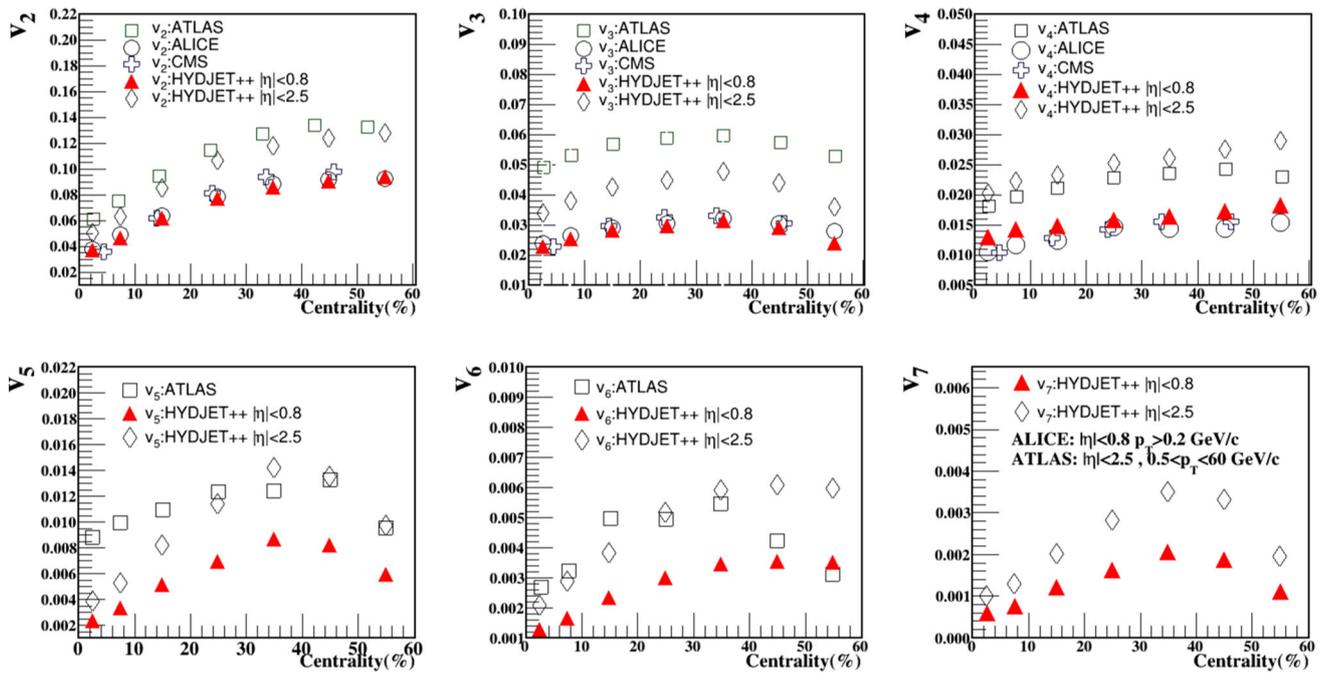

**Fig. 2** Centrality dependence of calculated minimum bias $v_n$ ($n = 2$–$7$) for Xe–Xe collisions at 5.44 TeV center of mass energy along with experimental data [45–47] for comparison. The figure presents results in $|\eta|< 0.8$ with $p_T > 0.2$ GeV/c and $|\eta|< 2.5$ with $0.5 < p_T < 60$ GeV/c kinematic ranges for (0–60)% centrality range

Anisotropic flow coefficients present a significant dependence on centrality of collision (see Fig. 2). $v_n$ decreases as the order 'n' increases. As we move from most central to most peripheral collisions, flow increases and then decreases in most peripheral collisions. The effect of pseudorapidity cut is only visible in terms of quantity. Considering charged-hadron multiplicity (which decreases from most-central to most-peripheral class of collisions) as a representative for system-size (in context to centrality of collision), we conclude that anisotropic flow decreases as collision system-size increases [23]. We have presented our model results in $|\eta|< 0.8$ and $|\eta|< 2.5$ kinematic ranges comparing model results with LHC experimental data [45–47]. At small pseudorapidities ($|\eta|< 0.8$), HYDJET++ model results show a suitable match with available ALICE and CMS experimental data. However, at large values of $\eta$ ($|\eta|< 2.5$), only elliptic flow and to some extent $v_4$ shows some agreement with ATLAS data in most-central collisions, while $v_3$ and other higher Fourier orders underpredict experimental data irrespective of the collision centrality. Thus, HYDJET++ model presents a good description of ALICE and CMS experimental data. However, it does not show a suitable quantitative match with ATLAS experimental data as a function of collision centrality. This is because HYDJET++ model uses Bjorken boost-invariant hydrodynamics which works well for mid-rapidity region and not at higher forward and backward rapidities where Landau hydrodynamics works more appropriately. Elliptic flow rises with rising impact parameter, whereas triangular flow falls at comparatively lower impact parameter. So, $v_4$ and $v_6$ have behavior similar to $v_2$. But, $v_6$ also being individually contributed by $v_3$, sees a fall while $v_4$ doesn't (only depends on $v_2$). The strong rise in $v_5$, $v_6$, and $v_7$ is the effect of elliptic flow in them and fall is a consequence of $v_3$. The variation with centrality is maximum for $v_2$, attributed to the large change in the second-order eccentricity [49] from central to midcentral collisions. The role of fluctuations is not well expressed from the model. In a recent article [47], it is observed that the removal of biasness from dijets by including all reaction channels in PYTHIA enhances the even-order Fourier harmonics while suppressing the odd-order harmonics. Thus, it is strongly possible that the lower-orders $v_2$ and $v_3$ might affect the generation of higher-order flow harmonics when produced by the mode-mixing technique. To understand it further, we need to study the response of these higher-order flow coefficients to the lower ones. We shall discuss this later.

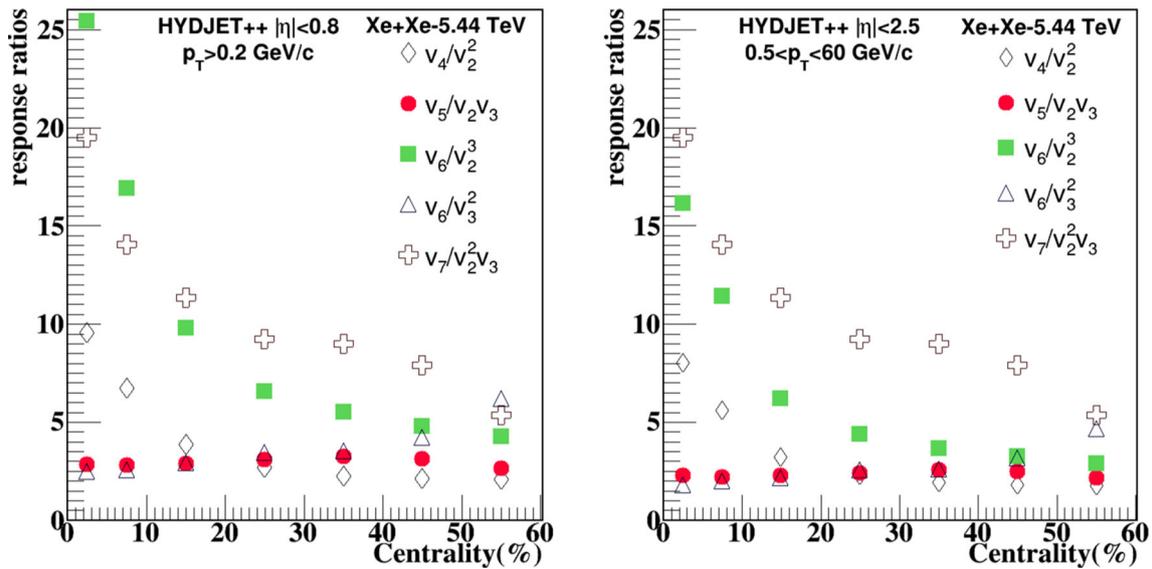

**Fig. 3** Response ratios of mean values of anisotropic flow coefficients $v_n$ ($n = 4$–$7$) as a function of collision centrality. The left panel of the figure presents results in $|\eta| < 0.8$ with $p_T > 0.2$ GeV/c, whereas the right panel shows results in $|\eta| < 2.5$ with $0.5 < p_T < 60$ GeV/c kinematic ranges for (0–60)% centrality range

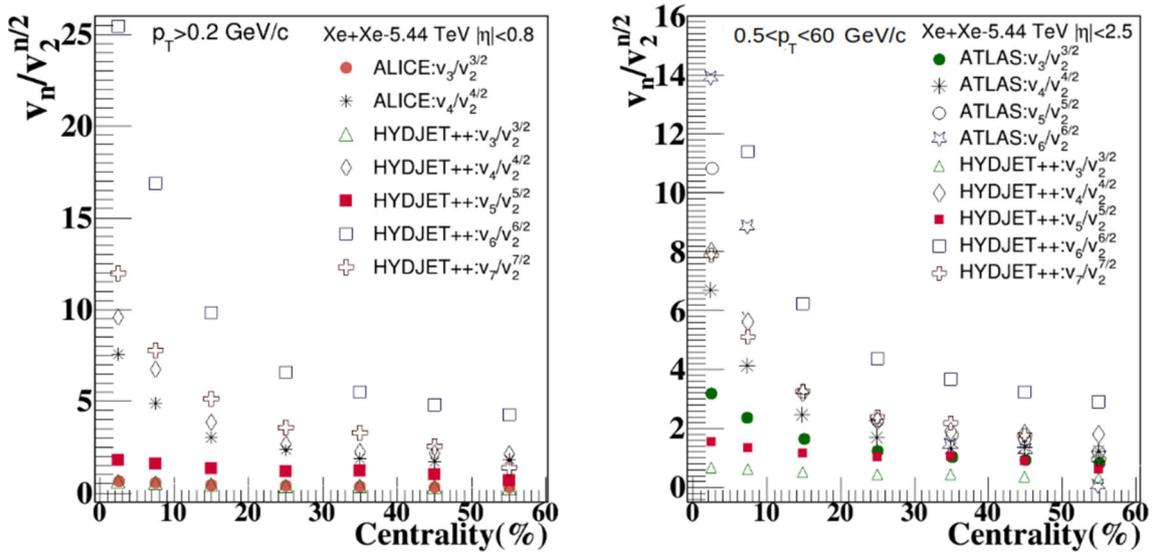

**Fig. 4** The centrality dependence of mean values of scaled harmonics $v_n/v_2^{n/2}$ ($n = 3$–$7$) in the (0–60)% collision centrality range. The left panel of the figure presents results in $|\eta| < 0.8$ with $p_T > 0.2$ GeV/c, whereas the right panel shows results in $|\eta| < 2.5$ with $0.5 < p_T < 60$ GeV/c kinematic ranges. HYDJET++ model results have been compared with ALICE [45] and ATLAS [47] experimental data

Elliptic flow of particles arises mainly due to the geometry of collision [15]. The evolution of $v_2$ with collision centrality is likely as it varies approximately linearly with the eccentricity of the overlap zone of the colliding nuclei. In case of peripheral collisions, elliptic flow generation is restricted by the shorter lifetime of the system formed upon collision, and due to the lower contribution of hadronic interactions and eccentricity fluctuations [50, 51]. Hence, such correlation is observed due to $v_2$. The $_{54}$Xe$^{129}$ nucleus is smaller as well as deformed contrary to the large spherical $_{82}$Pb$^{208}$ nucleus collided at LHC. Therefore, Xe–Xe collisions have larger event-by-event fluctuations in the initial geometry compared to Pb–Pb collisions [52]. At the same time, the produced QGP fireball in smaller Xe–Xe collision systems has larger viscous effects in its hydrodynamic expansion [52–54]. On the other hand, triangular flow $v_3$ is produced entirely by the initial participant fluctuations and is largely sensitive to the transport coefficients such as the shear viscosity of the created medium, which tends to abolish the azimuthal anisotropy, especially for smaller-size systems. This needs to be discussed more in detail and now it becomes necessary to investigate the behavior of the nonlinear part of anisotropic flow coefficients as a function of collision centrality. An important point to be noted here is that HYDJET++ deals with parameterized ideal hydrodynamics with no inclusion of any dissipative effects. Therefore, the transport properties of the QGP medium cannot be directly understood through this model framework, but we can speculate about these properties through the non-linear behavior of the inter-flow correlations through the response coefficients.

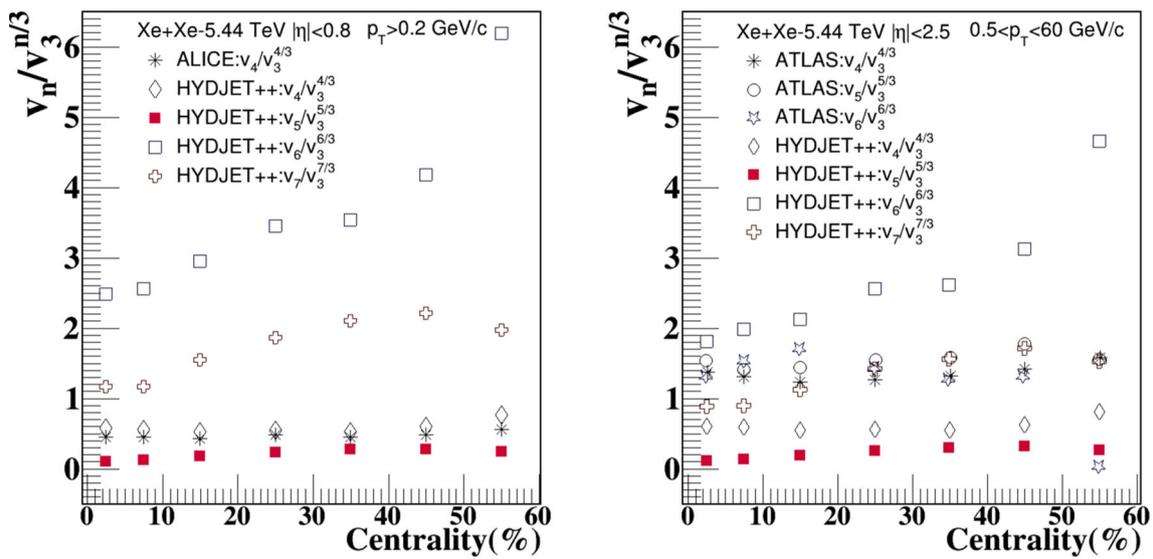

**Fig. 5** Centrality dependence of mean values of scaled harmonics $v_n/v_3^{n/3}$ ($n = 4$–7) for (0–60)% centrality range. The left panel of the figure presents results in $|\eta| < 0.8$ with $p_T > 0.2$ GeV/c, whereas the right panel shows results in $|\eta| < 2.5$ with $0.5 < p_T < 60$ GeV/c kinematic ranges. HYDJET++ model results have been compared with ALICE [45] and ATLAS [47] experimental data

Figure 3 shows the response ratios of minimum bias higher-order flow harmonics $v_n$ ($n = 4$–7) to $v_2$ and $v_3$ as a function of collision centrality obtained from Monte Carlo HYDJET++ model in two different kinematic ranges. The left panel presents response ratios for $|\eta| < 0.8$ with $p_T > 0.2$ GeV/c, while the right panel depicts response ratios for $|\eta| < 2.5$ with $0.5 < p_T < 60$ GeV/c kinematic range. The results show that the hydrodynamics holds the magnitude as well as the sign of the nonlinear response of the flow coefficients. In other words, we can say that, as we move from most-central to most-peripheral class of collisions, the viscosity of the medium increases or the hydrodynamics makes a transition from ideal to viscous. The proper response coefficients are independent of the initial density profile. So, they should not depend on collision centrality. However, we observe the response ratios ($v_4$, $v_6 − v_2$, $v_7$) in both pseudorapidity regions dependent on collision centrality. This dependence observed, largely comes from the strong dependence of the elliptic flow coefficient on collision centrality. Also, there may be some (non-negligible) contributions from other mode-mixing terms mentioned in the equations from (12)–(18) in article [20], which we have not included in the calculation of higher-order flow harmonics. The response ratios are smaller in magnitude at higher pseudorapidities ($|\eta| < 2.5$) compared to those at smaller pseudorapidity range ($|\eta| < 0.8$). Higher flow harmonics have similar behavior as in response to elliptic flow. The ratio decreases with collision centrality. $v_4$, $v_6 − v_2$, $v_7$ response ratios decrease with increasing impact parameter, whereas $v_6 − v_3$ response ratio increases with increasing impact parameter. Nonlinear response coefficients are persistent to freeze-out [16] but least understood in hydrodynamic calculations. They do not directly depend on the initial density profile but can be strongly discerned at the freeze-out temperature [55]. In HYDJET++, the produced fireball has geometrical irregularities in different directions of phase space at the time of freeze-out, where the QCD medium produced is assumed to evolve according to the Bjorken boost-invariant hydrodynamics. These irregularities are related to the initial spatial distribution of the participating nucleons in the collision region. Further deep studies are required to pin down the sensitivity of response coefficients to model parameters.

To have a more clear picture, we need to explicitly visualize (dependence on $v_2$ and $v_3$ separately) the response of the nonlinear part of flow harmonics as a function of collision centrality (see Figs. 4 and 5). Here, we employ the $v_n/v_m^{n/m}$ ratio technique to check the scaling behavior of the flow coefficients. We have presented results for two pseudorapidity ranges, where the left panel shows results for $|\eta| < 0.8$ and $p_T > 0.2$ GeV/c whereas the right panel shows results in $|\eta| < 2.5$ and $0.5 < p_T < 60$ GeV/c kinematic ranges for (0–60)% centrality range. It is seen that the results are higher for lower $|\eta|$ values and lower at high pseudorapidity ranges. The scaling depends on the harmonic order 'n' and the centrality of collision. The average values of $v_n/v_2^{n/2}$ (see Fig. 4) are larger in central collisions and become smaller (almost approaching zero) in peripheral collisions. HYDJET++ model results present a suitable match with ALICE experimental data. However, such match is not achieved with ATLAS experiment. This discrepancy may be attributed to the fact that HYDJET++ uses Bjorken boost-invariant hydrodynamics which is not defined at larger rapidities. Therefore, the model inferences may get altered if a proper hydrodynamical treatment is incorporated in HYDJET++ at large rapidities. The scaling order is very much clear over 0–30% centrality interval where $v_6/v_2^{6/2}$ is maximum whereas $v_3/v_2^{3/2}$ is observed to be minimum. However, beyond 30% centrality, the difference between the ratios decreases. Similar observations are made when the scaling is performed with $v_3$ (see Fig. 5). The scaling ratios have (albeit-) similar values for different 'n.' The dependence of hydrodynamics on the magnitude as well as the sign of the nonlinear response of the flow coefficients discussed in Fig. 3 is clearly seen here arising due to the presence of $v_3$. One of the reasons for the difference in the scaling by $v_2$ and $v_3$ in Figs. 4 and 5, respectively, is because, $v_2$ is strongly dependent on collision centrality and $v_3$ comparatively has a weaker centrality dependence. The scaling or correlation of individual response toward $v_2$ and $v_3$ is quite opposite. This means that moving from most-central to most-peripheral collisions, the hydrodynamics of the system shifts from (less viscous) ideal to (more) viscous whereas $v_3$ behavior is opposite. It shifts the system hydrodynamics lesser viscous or remains ideal. However, due to the strength of $v_2$ being effectively more than $v_3$, it is observed that all over, the system hydrodynamics makes a transition from ideal to a more viscous one.

## 4 Summary and outlook

Higher-order anisotropic flow coefficients $v_n$ (4–7) as a function of collision centrality are produced as well as presented for deformed Xe–Xe collisions at 5.44 TeV using HYDJET++ model in two kinematic regions; first, the $|\eta| < 0.8$ with $p_T > 0.2$ GeV/c, and second $|\eta| < 2.5$ with $0.5 < p_T < 60$ GeV/c, for $0 < p_T < 5$ GeV/c kinematic ranges, similar to the available experimental data. The gist of our findings is as follows:

1. Higher harmonic anisotropic flow orders $v_n$(4–7) are generated by the interference of elliptic and triangular flows. The results are compared with recent ALICE, ATLAS and CMS experimental data at LHC. Higher orders in anisotropic flow are centrality dependent. Flow decreases as collision system-size and harmonic order 'n' increases. HYDJET++ model results show a suitable match with ALICE and CMS experimental data but fail to explain ATLAS experimental results at higher pseudorapidity range. $v_2$ is strongly centrality dependent whereas $v_3$ is comparatively weaker. So, the strong dependence in higher flow coefficients is due to elliptic flow, while the weaker dependence is attributed to centrality dependence of $v_3$.
2. The response ratios of minimum bias higher-order flow harmonics to $v_2$ and $v_3$ as a function of collision centrality. Hydrodynamics is responsible for the magnitude as well as the sign of the nonlinear response of the flow coefficients. The response ratios are centrality dependent and smaller at higher pseudorapidities. The dependence is attributed to the linear part of the flow coefficients. Higher flow harmonics share similar behavior in response to $v_2$. The ratio decreases with collision centrality. The ratio is weakly centrality dependent in respect to $v_3$ and increases with collision centrality. Nonlinear response coefficients are persistent to freeze-out and negligibly depend on initial density profile but are strongly understood at the freeze-out temperature where Bjorken boost-invariant hydrodynamics plays an important role.
3. We studied the individual response of higher-order flow coefficients to the lower-order flow coefficients through a power-law (relation $v_n/v_m^{n/m}$) scaling technique as a function of collision centrality. The scaling depends on the harmonic order 'n' and the centrality of collision. It is higher for lower $|\eta|$ values and lower at higher $|\eta|$ ranges. The scaling order is very much clear over 0–30% centrality interval where $v_6/v_2^{6/2}$ is maximum whereas $v_3/v_2^{3/2}$ is minimum. Beyond 30% centrality, the difference in the ratios decreases. Similar observations are made when the scaling is performed with $v_3$. The difference in the scaling by $v_2$ and $v_3$ is attributed to the fact that $v_2$ is strongly dependent on collision centrality compared to $v_3$ which has weaker centrality dependence. The hydrodynamics of the smaller deformed Xe–Xe collision systems shifts toward more viscous one, as we move from most-central to most peripheral class of collisions.

**Acknowledgements** BKS sincerely acknowledges financial support from the Institutions of Eminence (IoE) BHU grant number-6031. SP acknowledges the financial support obtained from UGC under research fellowship scheme during the work.

**Data Availability Statement** This manuscript has associated data or the data will not be deposited. [Authors' comment: This manuscript has associated experimental data provided in the references data repository publicly. All data included in the manuscript from HYDJET++ model will not be deposited but are available upon request from the authors].